# A Progressive Network Management Architecture Enabled By Java Technology

**Damianos Gavalas**

Communication Networks Research Group

Electronic Systems Engineering Dept., University of Essex, Colchester, CO4 3SQ, U.K.

Tel: +44 1206 872425

dgaval@essex.ac.uk

**Dominic Greenwood**

Fujitsu Telecoms. Europe Research Group

Northgate House, St. Peters Street, Colchester, CO1 1HH, U.K.

Tel: +44 1206 363002

D.Greenwood@ftel.co.uk

**Mohammed Ghanbari**

Communication Networks Research Group

Electronic Systems Engineering Dept., University of Essex, Colchester, CO4 3SQ, U.K.

Tel: +44 1206 872434

ghan@essex.ac.uk

**Mike O'Mahony**

Communication Networks Research Group

Electronic Systems Engineering Dept., University of Essex, Colchester, CO4 3SQ, U.K.

Tel: +44 1206 872277

mikej@essex.ac.uk

**Abstract**

This paper proposes a framework based completely on Java technology. The advantages brought about by the use of Java in network management answer some critical problems existing in current systems. With this work we address several factors concerning interoperability and security in heterogeneous network environments. Specifically, we present a manager application and a multithreaded agent engine that make use of a lightweight communication mechanism for message exchange. A MIB parser is introduced to accelerate handling of incoming management requests, and the RSA public-key cryptosystem is implemented to provide both encryption and authentication features. Results, measured in terms of response time, compare favourably with other published work and standard management frameworks.

## 1. INTRODUCTION

The explosion in the size and complexity of today's local and wide area networks, combined with the increasing demands placed upon them for their resources has resulted in establishing Network Management (NM) as a factor of paramount importance. To further complicate matters, the concept of a single vendor network has vanished into history. Even if a company's network is provided by a single vendor, very soon the requirement to interconnect with related companies by using a mixture of public and/or private networks make the single vendor objective unrealistic. Therefore, an important goal of NM is to support the heterogeneous integrated environment of a network that contains multi-vendor hosts, software packages and carriers.

The current state of the art in NM involves a management application (manager) and the managed entities (agents), embedded within Network Elements (NEs). Management interactions make use of the Client/Server model, with the manager collecting status data and setting control variables through the agents. The communication between the managing and the managed entities is facilitated by NM protocols such as the Simple Network Management Protocol (SNMP) [2], part of the TCP/IP protocol suite and the Common Management Information Protocol (CMIP) [3] used in public telecommunication networks. Within these protocols, abstractions of physical resources in a network are represented by managed objects. Collections of managed objects are grouped into a tree-structured Management Information Base (MIB) or Management Information Tree (MIT) following the Abstract Syntax Notation 1 (ASN.1) format and encoded for transport using the BER encoding [4]. The Internet standard MIB-II [6] is an example of a MIB being supported by all the SNMP-managed NEs. Of course, the implementation of the agents (which are typically 'closed', non extensible applications) are left to the vendors, as long as they understand the protocol standard 'language' and follow the MIB or MIT format and encoding definitions.

Despite its wide use, SNMP is known to have several significant disadvantages. The most important among these is the poor community strings-based authentication pattern, adopted by both SNMPv1 and SNMPv2c [9], in addition to the lack of encryption of sensitive NM information. The issue of portability between different hardware architectures or operating systems and the difficulty to extend agent functionality should be also stressed. Another limitation of SNMP is related to the use of BER encoding, which is renowned for its inefficiency. A second issue, regarding the protocol's efficiency is within SNMP itself; SNMP *varbind* lists are relatively expensive, because the Object Identifiers (OIDs) used to name variables usually take much more space than the actual values. Also, the absence of an efficient table retrieval mechanism means that the overall protocol efficiency suffers from repeated message exchanges (and repeated computations on the agent side).

Java technology has recently attracted tremendous attention in the NM field both in research [5][7] and at a commercial level [14]. Both [5] and [7] use TCP as transport protocol, and give a solution to the portability but not to the security problems of SNMP. In addition, they adopt a heavyweight communication mechanism that causes a serious impact on the Network Management System (NMS) performance, both in terms of bandwidth usage and response time.

The objective of this work is to explore the feasibility of integrating Java technology in NM. The Java approach explained in this paper is intended to overcome some of the SNMP limitations and reduce the complexity involved in NM. Two different aspects of NM that exhibit several attractive features in comparison with the SNMP-based management are considered. The first involves the manager side; we present an application in the form of a browser tool that automatically discovers network topology and interacts with the managed NEs using an efficient and lightweight communication scheme. The second aspect involves a portable and extensible agent engine that, due to its multithreaded nature, supports concurrent queries from one or more managers. In addition, an efficient MIB parsing method has been implemented at the agent side to speed incoming manager request processing. Security issues have also been investigated such that manager requests are authenticated while data returned from the agents are encrypted. Also, either TCP or UDP protocol stacks can be used for message exchange. As a whole, the presented framework consists of 38 classes amounting to no more than 250 Kb. This relatively small footprint makes the method even more attractive for network devices with limited storage capacity.

The rest of the paper is organised as follows: Section 2 summarises the characteristics of Java that make it a suitable platform for the development of NM applications. Section 3 provides an overview of the MIB parser, the manager and the agent application, and describes the exchanged messages between the managing and the managed entities. This section also highlights the differences with the approach taken in [5]. Section 4 deals with associated security issues. Experimental results are reported and discussed in Section 5 whilst Section 6 draws conclusions and considers directions for future work.

## 2. WHY JAVA FOR NETWORK MANAGEMENT?

Network computing based on Java technology is an emerging technology that provides the basis for addressing limitations with legacy management systems. Java can be considered a technology rather than merely as another programming language due to its 'standard' implementation that includes an industrially supported network management infrastructure (JMAPI) [12].

The proposed infrastructure has been entirely developed in Java, chosen due to its numerous attractive features. Firstly, it is architecturally neutral and hence offers the platform independence (portability) required for the management of a multi-faceted heterogeneous environment. Furthermore, it has strong networking support through multithreading, allowing the concurrent execution of several processes. Java also has the facility to obtain local system information through native methods. In addition, the use of management applets allows an operator to remotely manage a network through a standard web browser. The fundamental assumption for our framework is the presence of a Java Virtual Machine (JVM) in every managed NE. Current trends [11] indicate that JVMs may soon be integrated into many network and computing resources.

Our framework has been developed using the Java Developer Kit (JDK) 1.1.7, with the Swing package of JDK 1.2, Beta 4, used for the construction of the user interfaces.

## 3. NM ARCHITECTURE IMPLEMENTATION ISSUES

The main components of our proposed architecture consist of a Manager Browser and an Agent Application. The manager browser is responsible for monitoring NEs by sending SNMP-like requests to perform management tasks. The agent application runs as a daemon on each of the managed devices providing an interface between the NMS and the NE resources and serving manager requests.

In addition, a MIB parser constructs a Random Access File (RAF) used by the agent to build a tree representation of the MIB objects. The use of the RAF structure reduces the time taken by the agent to map a requested OID to the corresponding object name. The implementation has been tested using the MIB-II definition [6], although the parser accepts any MIB format.

### 3.1. The Manager Application

The manager is a Java application that executes monitoring and control operations through its interaction with the agent processes. The manager adopts a browser style Graphical User Interface (GUI) to ease user interaction.

#### 3.1.1. Browser Overview

The GUI, shown in Figure 1, is the application tool through which the operator controls the managed environment. It facilitates selection of the managed devices to be queried, the transport protocol (TCP or UDP) and settings for both the OID and community strings of the requested MIB object. In addition, the end user can determine (in runtime) whether the messages exchanged are being authenticated and the management data encrypted or not.

The browser also maintains a list of the agents, which initially includes the host names of all agent processes in operation at manager initialisation. These active processes are 'discovered' via a manager broadcast poll. Whenever a new agent process starts operation after manager start-up, the manager is notified and the host name appended to the list. Similarly, whenever an agent process terminates, it notifies the manager application, which in turn removes the host name from the browser's list.

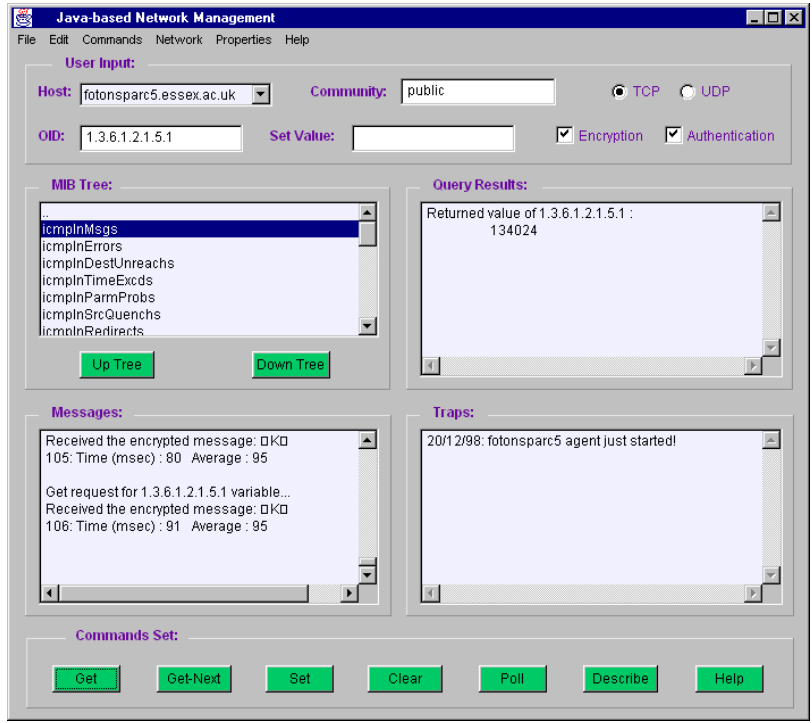

**Figure 1: The Manager Graphical User Interface (GUI)**

Additional facilities allow the operator to:

- Navigate the MIB tree;
- Issue an 'SNMP-like' request;
- Release a connection;
- Initiate an automated network discovery process;
- Obtain an on-line description of a MIB variable (see Figure 2);
- Set/modify the polling frequency, in runtime;
- Log operational results into a log file.

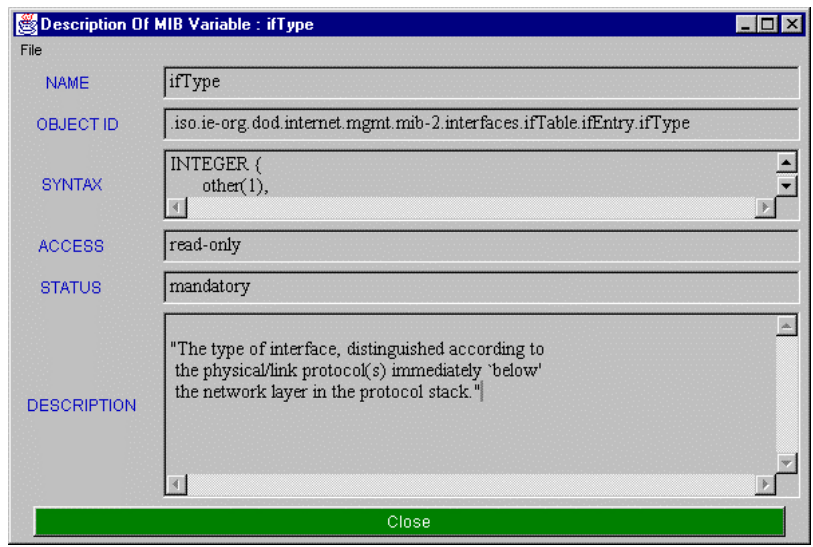

**Figure 2: On-line description of a MIB variable**

### 3.1.2. Manager Requests

As has been already mentioned, in our approach either TCP or UDP transport protocol may be used for conveying the exchanged requests/responses. The issue of whether management data should be carried over a reliable protocol such as TCP or an unreliable protocol such as UDP has been debated for years. We believe that the network administration should have the choice. Clearly, it is the duty of the administrator to ensure that the proportion of the management data remains low compared to the user data. Thus, in case of oversized networks where packet losses occur due to bursty traffic, management data should be transported with a reliable protocol. TCP should also be the protocol choice in cases that the number of retrieved values per managed device increases (as the number of transferred messages per connection increases, the efficiency of TCP improves). Conversely, for expensive intercontinental WAN links, management data are far less important than user traffic; in these cases management data should be carried over an unreliable transport protocol such as UDP.

A simple management query involves establishing a TCP or UDP (depending on the protocol stack choice) socket connection, which remains open for the whole duration of the transaction. Messaging will be then carried by TCP segments or UDP datagrams, respectively. This lightweight communication mechanism differs greatly from that adopted in [5], which relies on Java classes to pass information through the network. In particular, with reference to [5], these classes include the requested OID string and have to be compiled before being transmitted. The receiving agent loads the received class and retrieves the OID string. The same scheme is also used in the return direction. Still, while adopting a heavyweight messaging approach (unnecessary transfers of classes) this framework does not overcome the security weaknesses of SNMP, as the requests are not authenticated and the content of the transferred classes is 'readable'. This process also exhibits the drawback of delay imposed by time-intensive compilation and class loading.

In our proposed framework, the following request types may be sent from the manager to the agent entity:

- 'Initialise' request: This follows a successful connection establishment. The agent responds by returning the first level of the MIB tree.

- 'Next_Level' / 'Upper_Level' requests: These facilitate the navigation to lower or upper levels of the MIB tree, respectively. In the case of a 'Next_Level' request, the children (next level nodes) of the currently selected node in the browser's MIB tree list will be returned. Otherwise, the agent returns the list of the upper level nodes.

- 'Get' / 'Get-Next' / 'Set' requests: These have the same functionality as their SNMP counterparts. Their invocation causes the manager to send, together with the command string, the OID string it refers to. The OID string is checked before transmission to ensure its validity.

- 'Describe' request: This provides on-line viewing of the information (textual OID, syntax, access rights, status, description, etc.) associated with the selected MIB node.

- 'Connection_Release' request: When received by the agent, the socket connection is released and the separate process (thread) that was created to carry out the manager's requests is terminated.

An alternative option for socket-based communication would be the use of Java Remote Method Invocation (RMI). Like sockets, RMI offers a bi-directional association: once an RMI client has bound to an RMI server, both of them can send data to each other. RMI is an elegant solution in terms of design, as it gives a fully object-oriented view of network management while it eases the design of complex applications. However, current RMI implementations are considerably slower and tend to have a high demand on system and network resources, which indicates that RMI-based NM is not sufficiently scalable for widespread applications. The same applies to the Object Management Group's (OMG) Common Object Request Broker Architecture (CORBA); despite the language independence that it brings, CORBA is a complex technology that makes use of heavyweight synchronous protocols, incurring network transport overheads. Thus, for the time being, sockets offer a preferable communication mechanism for NM.

### 3.2. MIB Parser

The MIB parser is a standalone Java application, executed prior to agent initialisation that creates a RAF (through the *java.io.RandomAccessFile* class) using a standard MIB text file for input. The RAF (Table 1) consists of a list of records, one for each of the objects contained in the MIB file, and by its nature allows immediate access to specific records. During agent initialisation, the RAF is used to build a MIB tree representation that is stored in system memory and enhances agent efficiency when handling manager requests. The usefulness of both the RAF and the MIB tree will be clearly shown in the next section.

| Object Name | mib-2 | System | Interfaces | ..... | IfType | ..... | snmpEnable AuthenTraps |
|---|---|---|---|---|---|---|---|
| Syntax |  |  |  |  | INTEGER .. |  | INTEGER .. |
| Access |  |  |  |  | read-only |  | read-only |
| Status |  |  |  |  | Mandatory |  | mandatory |
| Description |  |  |  |  | "The type... stack" |  | "Indicates ... System" |
| Parent Node | Mgmt | mib-2 | mib-2 |  | IfEntry |  | snmp |
| Identifier | 1 | 1 | 2 |  | 3 |  | 30 |
| *Record Index* | *0* | *1* | *2* |  | *23* |  | *200* |

**Table 1: Random Access File structure**

The parser application compares each of the parsed strings with selected keywords. The MIB file comments are skipped in order to speed the parsing procedure. Each record, included in the resulting RAF contains object name, syntax, access, status and description fields.

### 3.3. The Java Agent Engine

The Java Agent is an application that runs as a background process on a managed node, listening on well-known TCP and UDP ports and waiting for a manager connection request. Upon reception of such a request, the main agent thread forks a slave thread that manages all the requests sent by the manager for the duration of the connection (see Figure 3). The main agent thread is then freed and returns to listening mode awaiting new connection requests. In this manner, a single agent can serve multiple managers simultaneously. When the release of a connection is requested, the slave thread terminates.

The Net Discovery thread is used to assist the manager to discover the active agent processes. It listens on a pre-defined UDP port waiting for manager broadcast messages. When receiving such a message this thread replies by sending the manager the IP address of the agent host.

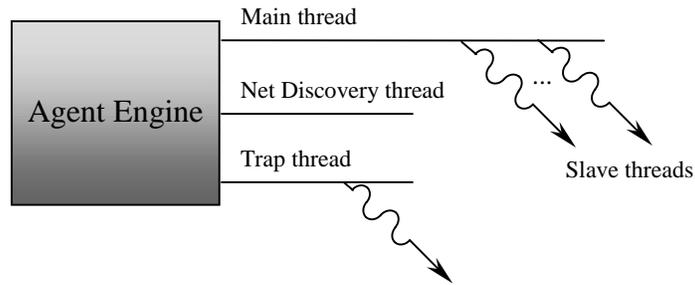

**Figure 3: Threads of the Agent engine**

### 3.4. MIB Tree Structure Description

During initialisation, the agent opens the RAF created by the MIB parser and sequentially reads the records stored within. For each of these, a node is inserted into a tree structure (MIB tree) created in system memory. Each of these nodes consists of:

1. The name of the object that the node represents.
2. A vector (dynamic array) containing the next level nodes.
3. An index number indicating the location (offset) of the corresponding record in the RAF.

The agent application operation is summarised by the flow diagram in Figure 4.

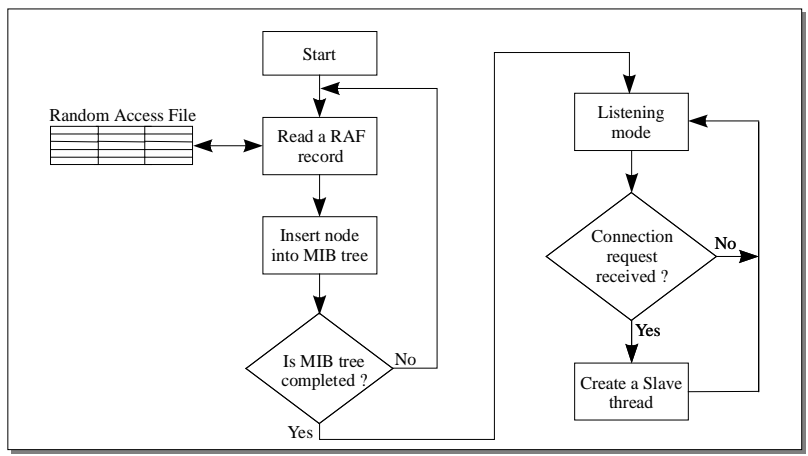

**Figure 4: Agent operation flow diagram**

Insertion of a node into the MIB tree is achieved through a recursive procedure. The agent makes a search on the existing MIB tree structure, recursively reaching the parent node. If, for instance, the MIB-II object 'ifType' is to be inserted, 'ifEntry' will be its parent node. When the parent node is located, a new tree node is constructed containing the name of the inserted object, an empty vector and the index to the RAF, while the vector of the parent node (the list of its 'children') is updated. Figure 5 illustrates the resulting mapping of the MIB tree nodes to the RAF record structure.

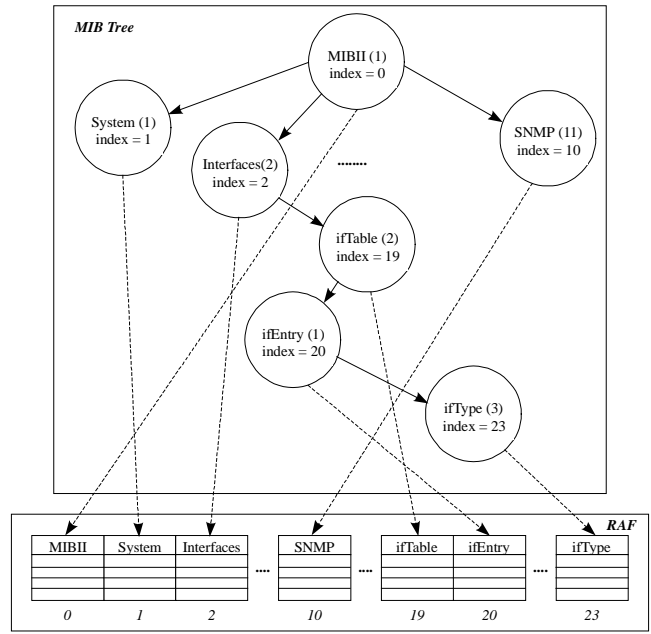

**Figure 5: Mapping of the MIB tree structure to the Random Access File**

### 3.5. Handling of Manager Requests

The existence of the MIB tree greatly reduces the agent's response time to manager requests. Thus, the manager requests, mentioned in section 3.1.2, are handled as indicated below:

- 'Initialise' request: The root of the MIB tree is returned.
- 'Next_Level' / 'Upper_Level' requests: A search method is invoked that traces the node for which the next or the upper tree level is requested. In the former case, the agent returns the located node's vector (containing the next level nodes), whereas in the latter the vector of that node's 'grandparent' is returned.
- 'Get' request: The agent parses the received OID string and follows the corresponding path in the MIB tree. If the object requested belongs to the system MIB-II group, its value is immediately retrieved; otherwise the system's kernel is accessed through native functions. 'Get-Next' and 'Set' requests are handled similarly.
- 'Describe' request: In this case, when the requested MIB tree node is located the offset of the associated record is also read. The record is then directly accessed and its fields returned by separate TCP segments (or UDP datagrams) to the manager. Figure 6 demonstrates how the slave thread manages the 'Get' and 'Describe' requests.
- 'Connection_Release' request: The socket connection closes and the slave thread dies.

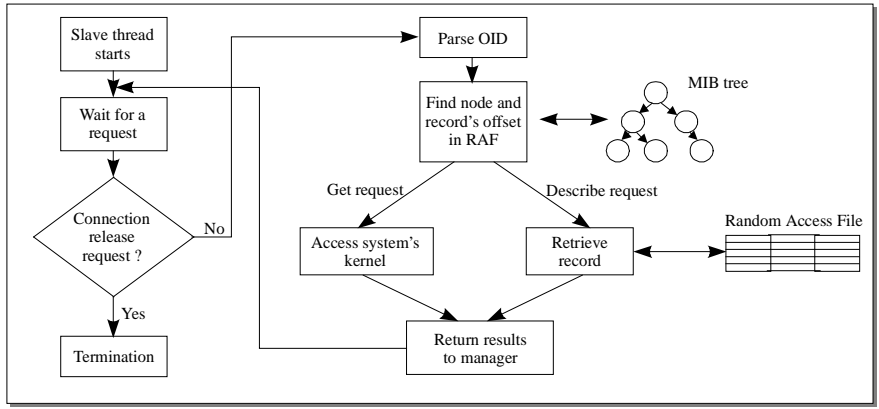

**Figure 6: Slave thread operation for 'Get' and 'Describe' request**

The use of the MIB tree structure represents a notable improvement in comparison with the parsing method described in [5]. Namely, according to the description given in [5] the mapping of an OID string to the corresponding object name is achieved through the parsing of the actual MIB text file. Hence, the parsing speed is dependent on the location of the requested object in the file, and a significant delay may be imposed.

### 3.5.1. Kernel Access Mechanism

The native methods available in Java are used to obtain NM-related system information. These methods are integrated into the Java code and are able to call functions written in other programming languages (C, in our case), suitable for accessing the local operating system. Integration is achieved through the Java Native Interface (JNI) [1]. The C code executes a lookup (based on streams, in the case of Solaris 2.5), and acquires a data structure containing all the objects of the MIB group of the requested variable. The object value of interest is then returned to the main Java program. The MIB-II groups currently implemented are ip, tcp, icmp and udp.

### 3.5.2. Traps Implementation

The agent application has been extended with event report capabilities through the implementation of a trap thread running in parallel with the main agent thread (see Figure 3). The primary task of this thread is to monitor several MIB objects by checking, at regular intervals, their values against a threshold defined by the manager. The trap thread uses the same mechanism described in the previous section to access the kernel (by forking a slave thread). In the case of a monitored object value exceeding the given threshold, the manager is notified by an event report. The manager application is able to append a new object to the list of those already being monitored by sending the name of the object together with a threshold value and the monitoring frequency.

## 4. SECURITY ISSUES

The proposed NMS is designed with several security features that overcome the apparent weaknesses of SNMP and ensure privacy whilst protecting from unauthorised access. In particular, the RSA public-key cryptosystem [8] has been implemented providing both encryption and authentication. The Java Security API (built on top of JDK 1.1) has not been chosen, due to the lack of encryption features, while the recently released Java

Cryptography Extension (JCE) 1.2 [13], which supplements the JDK 1.2, provides just a framework but not the actual implementation of encryption algorithms.

The RSA algorithm is based on the 'public-private pair of keys' paradigm and its security strength is related to the assumption that detection of the private key (needed to sign or decrypt the transmitted data) is highly improbable. Thus, the steps followed during a single transaction are:

i) The manager application signs the request using the private key. The digital signature generated is then sent to the Java agent along with the request message.

ii) The SNMP security scheme is maintained and hence the request community name is initially compared with the current agent's community name. Upon success, the agent executes an authentication function in conjunction with the public key and verifies whether the original request message is recovered (this ensures data integrity). If the second security level is successfully passed, the requested service is initiated else a message is returned to the originating host informing of the access denial to system information.

iii) The agent encrypts the data to be returned using the public key.

iv) The manager decrypts the data received by executing the decryption function, which makes use of the private key.

## 5. EXPERIMENTAL RESULTS AND DISCUSSION

Our experimental setup involves several machines connected through 10Mbps Ethernet segments. Specifically, a Solaris 2.5 UltraSparc-1, two Solaris 2.5.1 SparcStation4's and two 120MHz Pentium PCs running WinNT have been used.

The proposed infrastructure has been compared, in terms of response time, with the work of Luderer et al. [5], in addition to a standard SNMP package offered by University of California at Davis [15] (UCD-SNMP ver. 3.3.1), developed in C programming language. The application described in [5] has been implemented as a control model. The manager was run from the UltraSparc-1 and Java Agents were initialised on the other machines.

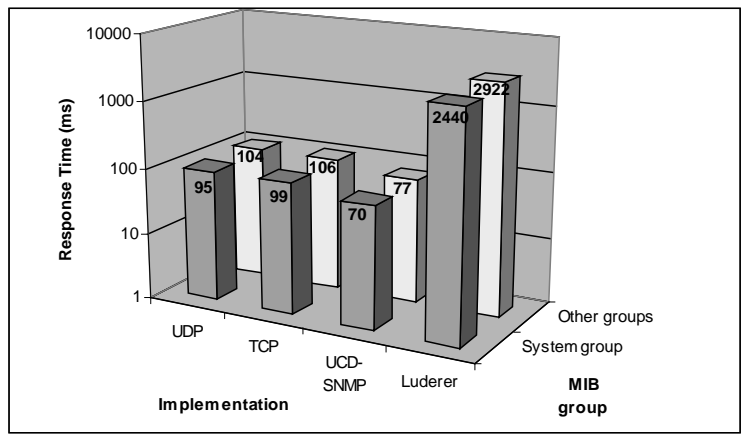

Figure 7: Comparison of response times measured for various implementations

The results shown in Figure 7 indicate that our proposed framework has a slightly longer response time than SNMP (due to the authentication and encryption procedures that are not present UCD-SNMP implementation), whilst an improvement in the order of 25 is achieved in comparison with [5]. Separate measurements have been taken for *system* and other MIB-II group variables (the latter require access to the system kernel in order to obtain the variable value). The response times were measured at the manager end and averaged for each group.

Clearly, thanks to the use of speed-up techniques such as the Just In Time (JIT) compilers, the poor execution speed of Java interpreted bytecode does not represent a significant problem any more.

## 6. CONCLUSIONS – FUTURE WORK

We have presented an NMS framework, including both a manager application and an agent engine, completely based on Java technology. Java has been found an excellent development platform, ideally suited to NM applications. Indeed, a number of intrinsic problems of NM can be easily addressed and brought closer to solution. These are: first, the portability across platforms or the independence from the underlying software and hardware architectures, and second, a set of security aspects, such as authentication of incoming requests and encryption of sensitive management information.

Moreover, the BER encoding used in the current SNMP frameworks is eliminated in this approach, since the interoperability is supported by the Java bytecode. The administrator is also given the option of using either the TCP or UDP protocol stacks depending on the managed network topology and the traffic conditions. A MIB parser tool improves the efficiency of the NMS, which seems to outperform a framework recently published in the literature.

Current and future work address the following areas:

- Optimisation of the manager and agent code in order to further decrease the response time.
- Construct an expanding and shrinking representation of the MIB tree on the manager browser.
- Employ mobile agent technology, so as to reduce the management data volume.